\documentclass[reprint,pra]{revtex4-1}
\usepackage{graphicx,amsmath,amsfonts}% Include figure files
\usepackage{dcolumn}% Align table columns on decimal point
\usepackage{xcolor}
\usepackage{bm}% bold math
\usepackage{enumerate}

\newcommand{\pV}{\partial V}
\newcommand{\n}{\hat{\mathbf{n}}}
\newcommand{\eps}{\boldsymbol{\varepsilon}}
\newcommand{\taup}{\left( \tau \right)}

\newcommand{\tp}{ \left( t \right) }
\newcommand{\fb}{ \mathbf{f} }
\newcommand{\hP}{\hat{\mathbf{P}}}
\renewcommand{\pl}{\hat{p}^\parallel}
\newcommand{\ql}{\hat{q}^\parallel}
\newcommand{\pt}{\hat{p}^\perp}
\newcommand{\qt}{\hat{q}^\perp}

\newcommand{\hPib}{\hat{\boldsymbol{\Pi}}}
\newcommand{\hPi}{\hat{\Pi}}

\newcommand{\Pib}{\boldsymbol{\Pi}}
\newcommand{\Pb}{\mathbf{P}}

\newcommand{\E}{\mathbf{E}}

\newcommand{\Ham}{\mathcal{H}}
\newcommand{\hHam}{\hat{\mathcal{H}}}
\newcommand{\Lgr}{\mathcal{L}}
\newcommand{\Vl}{\mathcal{V}^\parallel}
\newcommand{\Ml}{\mathcal{M}^\parallel}
\newcommand{\Vt}{\mathcal{V}^\perp}
\newcommand{\Mt}{\mathcal{M}^\perp}
\newcommand{\Ul}{\mathbf{U}^\parallel}
\newcommand{\Ut}{\mathbf{U}^\perp}
\newcommand{\Wl}{\mathbf{U}^\parallel}

\newcommand{\Ab}{\mathbf{A}}
\newcommand{\Cb}{\mathbf{C}}
\newcommand{\dAb}{\dot{\mathbf{A}}}

\newcommand{\xib}{\boldsymbol{\xi}}
\newcommand{\hxi}{\hat{\boldsymbol{\xi}}}
\newcommand{\hA}{\hat{A}}
\newcommand{\hAb}{\hat{\boldsymbol{A}}}

\newcommand{\I}{\tensor{\mathbf{I}}}

\newcommand{\dxib}{\dot{\boldsymbol{\xi}}}
\newcommand{\A}{\mathbf{A}}
\newcommand{\dA}{\dot{\mathbf{A}}}
\newcommand{\rbr}{\left( \mathbf{r} \right)}

\newcommand{\rb}{\mathbf{r}}
\newcommand{\rp}{\left( \mathbf{r} \right)}
\newcommand{\rpp}{\left( \mathbf{r}' \right)}
\newcommand{\dV}{\text{d}^3{\bf r}}
\newcommand{\dk}{\text{d}^3{\bf k}}
\newcommand{\dS}{\text{d}^2{\bf r}}

\newcommand{\BetaS}{\text{B}'}
\newcommand{\BetaSS}{\text{B}''}

\newcommand{\ha}{\hat{a}}
\newcommand{\hb}{\hat{b}}
\newcommand{\hc}{\hat{c}}

\begin{document}

%\preprint{APS/123-QED}

\title{Quantum Theory of \\ 
Radiative Decay Rate and Frequency Shift \\ of Surface Plasmon Modes \\}% Force line breaks with \\

\author{Carlo Forestiere}
\affiliation{ Department of Electrical Engineering and Information Technology, Universit\`{a} degli Studi di Napoli Federico II, via Claudio 21,
 Napoli, 80125, Italy}
\author{Giovanni Miano}
\affiliation{ Department of Electrical Engineering and Information Technology, Universit\`{a} degli Studi di Napoli Federico II, via Claudio 21,
 Napoli, 80125, Italy}
 \author{Mariano Pascale}
\affiliation{ Department of Electrical Engineering and Information Technology, Universit\`{a} degli Studi di Napoli Federico II, via Claudio 21,
 Napoli, 80125, Italy}
 \author{Roberto Tricarico}
\affiliation{ Department of Electrical Engineering and Information Technology, Universit\`{a} degli Studi di Napoli Federico II, via Claudio 21,
 Napoli, 80125, Italy}
 \affiliation{ICFO Institut de Ciències Fotòniques, The Barcelona Institute of Science and Technology, \\ 08860 Castelldefels, Barcelona, Spain}

\begin{abstract}
In this paper we study, in the time domain, the interaction between localized surface plasmons and photons in arbitrarily shaped metal nanoparticles,  by using the Hopfield approach to quantize the plasmon modes, where the electron oscillations are represented by a harmonic matter field linearly coupled to the electromagnetic radiation. The plasmon - photon coupling gives rise to dressed plasmon modes. We have found that the radiation does not induce a significant coupling among the different quasi-electrostatic plasmon modes for particles of size up to the plasma wavelength, but causes a frequency shift and an exponential decay in time of the modes. By solving the equations governing the expectation values of the plasmon creation and annihilation operators, we obtain a new closed-form full-wave expression for the decay rate and for the frequency shift of the plasmon modes. It is non-perturbative and it only depends on the surface charge distribution of the quasi-electrostatic plasmon modes. We validate the expression against the Mie theory for a nano-sphere of radius comparable to the plasma wavelength. Eventually, we investigate the decay rate and the frequency shift of the plasmon modes in isolated and interacting nanoparticle of non-canonical shape, as their size increases up to the plasma wavelength.
\end{abstract}

\maketitle

%\tableofcontents

\section{Introduction}

The interaction of light with collective oscillations of free electrons in metal nanoparticles, denoted as surface plasmons, is one of the most active branches of nano-optics \cite{novotny2012principles}: it enables the subwavelength confinement of electromagnetic fields \cite{koenderink2015nanophotonics} and the enhancement of light-matter interaction \cite{chikkaraddy2016single}. Even though many phenomena involving surface plasmons can be explained by either classical or semiclassical theories \cite{maier2007plasmonics}, plasmons have an inherent quantum nature \cite{tame2013quantum} which manifests through wave-particle duality, entanglement \cite{altewischer2002plasmon}, squeezing \cite{huck2009demonstration}, quantum interference \cite{PhysRevB.85.195463}, sub-Poissonian statistics \cite{tame2013quantum}, strong-coupling with single-molecules \cite{chikkaraddy2016single}. They are also a promising platform for implementing a quantum antenna \cite{Slepyan}.

The recent advances of femtosecond characterization techniques enabled the experimental investigation of the ultrafast plasmon dynamic in metal nanoparticles (e.g. \cite{lambrecht1997femtosecond,sonnichsen2002drastic,lehmann2000surface,anderson2010few,voisin2001ultrafast,muskens2006femtosecond,matsuo2001time,zavelani2015transient,pelton2006ultrafast,kokkinakis1972observation,Hartland:11}). In particular, the plasmon time-decay may occur either via the coupling with photons, i.e. the radiation damping, or via the decay into excitation of intraband or interband electron-hole pairs (e.g. \cite{sonnichsen2002drastic}), or via the collisions of the conduction electrons with the ion lattice. The majority of existing experimental studies have mainly focused on the plasmon dynamics in small particle regimes where the non-radiative decay processes are dominant. Nevertheless, as soon as the size of the particle becomes comparable with the resonant wavelength, the radiative damping becomes the dominant damping mechanism \cite{kreibig2013optical,Raschke2013}. 

The theoretical study of plasmon dynamics in the time-domain in terms of natural modes has been so far restricted to the realm of quasi-electrostatic approximation \cite{mayergoyz2007dynamics}, which is unable to describe the coupling of the surface plasmons with photons, unless specific perturbation/correction techniques are implemented \cite{waks2010cavity}. On the other side, frequency-domain techniques based on full-wave theories, such as the quasi-normal mode expansion \cite{lalanne2018light} or the Mie Theory \cite{kolwas2013damping}, have been used to retrieve synthetic parameters such as the resonance frequency and the radiative Q-factor.

In this paper, we study in the time domain the natural modes of surface plasmons in arbitrarily shaped nanoparticles by taking into account the coupling with the electromagnetic radiation, and we obtain new closed-form full-wave expressions for the decay rate and the frequency shift of the plasmon modes for nanoparticle sizes up to the plasma wavelength. The study is carried out in the framework of quantum theory following the seminal works of U. Fano \cite{fano1956atomic}, J. J. Hopfield \cite{hopfield1958theory}, and Barnett \cite{huttner1992dispersion,huttner1992quantization,huttner1991canonical}. 
 In the last few years, the quantization of surface plasmons in nanoparticles coupled to radiative electromagnetic field has been a very active research area \cite{vogel2006quantum,trugler2008strong,dung1998three,alpeggiani2014quantum,shishkov2016hermitian,bordo2019quantum}.
We describe the metal nanoparticle as an electron gas rigidly confined within the spatial domain occupied by the positive ion lattice. We model the coupling between the electron gas and the electromagnetic radiation by following Refs. \cite{fano1956atomic,hopfield1958theory,huttner1991canonical,huttner1992dispersion,huttner1992quantization}. In particular, we focus on the dynamics of the expectation values of the plasmon creation and annihilation operators in the ground state of the system. We represent
the transverse electromagnetic field in terms of plane waves. We expand the longitudinal component of the electron gas displacement field (matter field) and the longitudinal component of the electric field (Coulombian field) in terms of the quasi-electrostatic plasmon modes of the nanoparticle proposed by I. Mayergoyz \cite{fredkin2003resonant,mayergoyz2005electrostatic}.   As we shall see, this choice allows to  better understand  the physics of the problem, greatly simplifies its mathematical description, and considerably reduces the computational burden of its numerical solution. In particular, it allows the diagonalization of the plasmon contribution to the Hamiltonian of the system. The  plasmon - photon coupling gives rise to dressed plasmon modes. The induced coupling among the different quasi-electrostatic plasmon modes is weak for nanoparticles of sizes up to the plasma wavelength. The   plasmon - photon coupling  mainly causes both a frequency shift with respect to the quasi-electrostatic frequency, and an exponentially decay in time. 

In this paper we obtain a new closed-form expression for the frequency shift and for the decay rate in terms of the surface charge distribution of the quasi-electrostatic
plasmon modes. The formula is non-perturbative and can be easily numerically evaluated. We validate the formula for a sphere of size comparable to the plasmon frequency, against the Mie theory. Then, we study the decay rate and the frequency shift of plasmon modes in single nanoparticles and in dimers. We disregard non-radiative damping mechanisms because the aim of the paper is to provide a quantum description of the coupling between the plasmon oscillations and the photons, and its effects on the plasmon radiative decay rate and frequency shift.

The paper is organized as follows. In Sec. \ref{sec:LagHam}, we present the classical Lagrangian and Hamiltonian of the system “electron gas + electromagnetic field” in the Coulomb gauge. In Sec III we introduce a representation of the displacement vector field of the electron gas in terms of longitudinal and transverse components. In Sec. IV we first perform the canonical quantization by expressing the field operators in terms of the bases introduced in Sec. III, then we introduce the standard boson operators, and eventually write down the Heisenberg equations of the entire system. In Sec. V we first study the expectation value in the ground state of the plasmon creation and annihilation operators, and then, we obtain a new full-wave closed-form expression for the frequency shift and the decay rate by using the pole approximation technique. In Sec. VI we study the frequency shift and radiative decay of the plasmon natural modes of single nanoparticles and dimers with different sizes and shapes. We discuss the main results and conclude in Sec. VII. The main properties of the quasi-electrostatic plasmon modes are presented in Appendix A.

\section{Classical Lagrangian and Hamiltonian}
\label{sec:LagHam}
A metal nanoparticle occupies a region $V$ in free space: $\partial V$ is the boundary of $V$, $V_0$ is the vacuum region surrounding the nanoparticle, and $V_\infty = V \cup V_0 $ denotes the entire space. To describe the collective behavior of the conduction electrons we model them as an electron gas
confined within the positive ion lattice of the metal. The ion lattice is assumed to be rigid and the ion distribution is assumed to be uniform in $V$ with volumetric density $n_0$. At equilibrium, the distribution of the electron gas is assumed to be equal to $n_0$.

An electric field moves the electron gas. The motion is incompressible inside $V$ due to the uniformity of the electron gas distribution at equilibrium. As a consequence, in the linear regime, the volume charge density is zero inside $V$, but a surface charge with density  $\sigma \rp$ arises on $\partial V$ due to the charge conservation. The vector field  $\xib\rp$, defined in $V$, represents the displacement of the electron gas with respect to its equilibrium configuration. It is solenoidal in  $V$, but its normal component to $\partial V$ is different from zero. In the linear regime, the surface density $\sigma \rp$  is given by
\begin{equation}
    \sigma \rp = -e n_0 \xi_n \rp \quad \text{on}\, \partial V,
\end{equation}
and the current density field ${\bf j} \rp$ is given by
\begin{equation}
    \mathbf{j} \left( \mathbf{r} \right) = -e n_0 \dxib\rp \quad \text {in V,} \;
\end{equation}
where $\xi_n = \xib \cdot \n$, $\n$ denotes the normal to the surface pointing outward and $e$ is the absolute charge of the electrons; we  indicate  the partial derivative with respect to the time with a dot and we  omit  the time variable to simplify the notation. Since the displacement field is solenoidal in $V$, the total surface charge on $\partial V$  is equal to zero. The surface charge generates a restoring force that combined with the electron inertia gives rise to a continuum of harmonic oscillators, i.e., the surface plasmon oscillations. Since the aim of the paper is to study the effects of the surface plasmon-photon copupling on the natural motion of the system, we neglect the effects of the losses due to both the inter-band and the intraband transitions, and the collisions of the conduction electrons with the ion lattice. 

\subsection{System Lagrangian}

Following the standard approach of quantum electrodynamics \cite{cohen1997photons}, we start from the classical Lagrangian of the system “electron gas + electromagnetic field”. In the Coulomb gauge, the Lagrangian of the entire system is composed of three terms,  the  surface plasmon $\Lgr_p $, \ the  radiation field $\Lgr_{em}$, and  the  interaction term  $\Lgr_i$:
\begin{equation}
    \Lgr \left( \xib, \dxib, \A, \dA \right) = \Lgr_p + \Lgr_{em} + \Lgr_{i},
\end{equation}
where
\begin{subequations}
\begin{eqnarray}
     \nonumber
     \Lgr_p \left( \xib, \dxib \right) &=& \int_V \frac{\rho_0}{2} \dxib^2 \, \dV \\  &-& \int_{\pV} \int_{\pV} \frac{\rho_0 \omega_p^2}{2}\frac{\xi_n \rp \xi_n \rpp }{4 \pi\left| \mathbf{r} - \mathbf{r}' \right|} \dS \dS',\qquad \\
    \Lgr_{em} \left( \Ab, \dAb \right) &=& \int_{V_\infty} \left[ \frac{\varepsilon_0}{2} \dAb^2 - \frac{1}{2\mu_0} \left( \nabla \times \mathbf{A} \right)^2 \right]  \dV, \\
    \Lgr_{i} \left( \dxib, \Ab \right) &=& \int_{V} (-e n_0 \dxib)\cdot \Ab  \dV;
\end{eqnarray}
\end{subequations}
$\varepsilon_0$ is the dielectric constant of vacuum and $\mu_0$ is the magnetic permeability.  The vector field $\mathbf{A} \rp$  is the magnetic vector potential in the Coulomb gauge generated by the electron gas. The first contribution to $\Lgr_p$ is the kinetic energy of the electron gas, where $m_e^*$ is the effective mass of the electrons in the conduction band and $\rho_0 = n_0 m_e^*$  is the mass density of the electron gas at equilibrium. The second contribution to $\Lgr_p$ is the Coulomb potential energy due to the surface charges on  $\partial V$, where $\omega_p = \sqrt{ e^2 n_0 / \varepsilon_0 m_e^*}$ is the plasma frequency of the electron gas.
\subsection{System Hamiltonian}
We introduce the canonical variables:
\begin{subequations}
\begin{eqnarray}
    \Pb &=&  \frac{\delta{\Lgr}}{\delta\dot{\mathbf{\xib}}\rp}=\rho_0 \dot{\mathbf{\xib}} -e n_0 \mathbf{A} \quad \text{in} \; V, \\
    \Pib &=& \frac{\delta{\Lgr}}{\delta\dot{\mathbf{\A}}\rp}=\varepsilon_0 \dot{\bf A} \quad \text{in} \; V_\infty;
\end{eqnarray}
\end{subequations}
where $\delta{\Lgr}/\delta{\Cb}\rp$ is the functional derivative of $\Lgr$ with respect to the vector field $\Cb\rp$; $\Pb$ is the momentum density field canonically conjugated to the displacement field $\xib$, and the vector field $\Pib$ is canonically conjugated to the vector potential $\bf A$. The Hamiltonian of the system,
\begin{equation}
 \Ham = \int_V \dot{\xib }\cdot\Pb\dV+\int_{V_\infty} \dot{\Ab }\cdot\Pib\dV - \Lgr,
\end{equation}
is the sum of three terms: the kinetic energy $\Ham_k$, the Coulomb potential energy  $\Ham_{Coul}$, and the energy $\Ham_{em}$ of the radiation electromagnetic field,
\begin{equation}
\label{eq:H} 
    \Ham \left( \Pb, \xib, \Pib, \Ab \right)  = \Ham_k  + \Ham_{Coul}+ \Ham_{em} ,
\end{equation}
where
\begin{subequations}
\begin{eqnarray}
& \Ham_k \left( \Pb, \A \right) = \int_V \frac{1}{2\rho_0} \left( \Pb + e n_0 \A \right)^2 \dV,\\
& \Ham_{Coul} \left( \Pb, \Ab \right) = \int_{\partial V} \int_{\partial V}\frac{\rho_0 \omega_p^2}{2}\frac{\xi_n \rp \xi_n \rpp}{4\pi \left| \rb - \rb' \right|} \dS \dS',\\
& \Ham_{em} \left( \Pib, \Ab \right) = \int_{V_\infty} \left[ \frac{1}{2\varepsilon_0} \Pib^2 + \frac{1}{2\mu_0} \left( \nabla \times \Ab \right)^2 \right] \dV.
\end{eqnarray}
\end{subequations}
The Hamilton’s equations for the matter fields are:
\begin{subequations}
\begin{eqnarray}
    \label{eq:Xd}
    \dot{\xib }&=& \frac{\delta{\Ham}}{\delta{\mathbf{\Pb}}\rp} = \frac{1}{\rho_0} \left( {\bf P} + e n_0 {\bf A} \right) \quad \text{in} \, V, \\
    \label{eq:Pd}
    \dot{\Pb} &=& - \frac{\delta{\Ham}}{\delta{\mathbf{\xib}}\rp}=-n_0 e \E_{Coul} \quad \text{in} \, V,
\end{eqnarray}
\end{subequations}
where
\begin{equation}
    \E_{Coul} \rbr = - \nabla_{\bf r} \left[ \frac{1}{4\pi\varepsilon_0} \int_{\partial V} \frac{-e n_0 \xi_n \left( \rb '\right)}{\left| {\bf r} - {\bf r}' \right|} \right] \dS'
\end{equation}
is the contribution of the Coulomb term to the electric field generated by the electron gas. The Hamilton’s equations for the radiation fields are (in $V_\infty$ where {\bf j} = 0 in $V_0$):
\begin{subequations}
\begin{eqnarray}
    \label{eq:Adot} 
    \dot{\A} &=& \frac{\delta{\Ham}}{\delta{\mathbf{\Pib}}\rp} =\frac{1}{\varepsilon_0} \Pib \;, \\
    \label{eq:Pdot}
    \dot{\Pib} &=& -\frac{\delta{\Ham}}{\delta{\mathbf{\Ab}}\rp} = - \frac{1}{\mu_0} \nabla \times \nabla \times \A + \left(\mathbf {j} + \varepsilon_0 \dot{\mathbf{E}}_{Coul} \right).\qquad
\end{eqnarray}
\end{subequations}
The second term on the left hand side of Eq. \ref{eq:Pdot} (between round brackets) is the solenoidal component of the current density field in $V_\infty$.

\section{Helmholtz decomposition of the matter field}
We introduce the scalar product 
\begin{equation}
    \langle {\bf F}, {\bf D} \rangle_W = \int_W {\bf F}^* \rp \cdot \mathbf{D} \rp \dV,
\end{equation}
and the norm $\left\| {\bf F} \right\|_W = \sqrt{ \langle {\bf F}, {\bf F} \rangle_W }$.  If the integration domain is not explicitly indicated, the scalar product is defined over the volume $V$. 

Any sufficiently smooth solenoidal vector field $\mathbf{C}$ defined in the region $V$ can be resolved into the sum of two terms: i) a solenoidal and irrotational vector field $\mathbf{C}^\parallel$  with normal component on the boundary $\partial V$ different from zero; ii) a solenoidal and rotational (non zero curl) vector field $\Cb^\perp$  with normal component on the boundary $\partial V$ equal to zero. This is a particular case of the Helmholtz decomposition for vector fields defined on a bounded region. The vector fields $\mathbf{C}^\parallel$  and $\Cb^\perp$  are orthogonal according to the scalar product $\langle {\bf C}^\perp, {\bf C}^\parallel \rangle$.

The canonically conjugate vector fields $\Pb \rp$  and $\xib \rp$  are defined in $V$  where they are solenoidal, but their normal component to $\partial V$ is different from zero; instead $\Pib \rbr$ and $\Ab \rbr$ are solenoidal everywhere, their normal component to $\partial V$ is continuous. We now express the Hamiltonian in terms of the longitudinal and transverse components of $\Pb \rp$  and $\xib \rp$.  It is the sum of five terms: the plasmon term $\Ham_p$, the radiation field term $\Ham_{em}$, the plasmon - photon interaction term $\Ham_{i}$,  and the kinetic energy of the electron gas vibrating in the radiation field ($\Ham_{ii}'$ + $\Ham_{ii}''$),
\begin{subequations}
\begin{eqnarray}
  \;\nonumber \Ham_p &=& \int_V \frac{1}{2\rho_0} {\Pb^{\parallel}}^2  \dV \\
        &+& \int_{\partial V} \int_{\partial V} \frac{\rho_0 \omega_p^2}{2}\frac{\xi_n^\parallel  \rp \xi_n^\parallel \rpp}{4\pi \left| \rb - \rb' \right| }  \dV \dV', \\
  \Ham_{em} &=& \int_{V_\infty} \left[ \frac{1}{2 \varepsilon_0} \Pib^2 \dV +  \frac{1}{2 \mu_0} \left( \nabla \times {\bf A} \right)^2 \right] \dV,\quad \\
   \label{eq:appo}
   \Ham_{i} &=& \frac{e}{m_e^*} \int_V {\Pb^{\parallel}} \cdot {\bf A} \dV,
\end{eqnarray}
\end{subequations}
and
\begin{subequations}
\begin{eqnarray}
  \; \Ham_{ii}'  &=& \int_V \frac{1}{2\rho_0} \left( {\Pb^\perp}^2 + 2 e n_0 \Pb^ \perp \cdot \A \right) \dV, \\
   \label{eq:dia}
   \Ham_{ii}'' &=& \int_V \frac{e^2 n_0}{2 m_e^*} {\A}^2 \dV.
\end{eqnarray}
\end{subequations}
The contribution of the diamagnetic term $\Ham_{ii}''$ may be disregarded in a moderate coupling regime between the plasmons and photons as for systems of bound particles in the low-intensity radiation regime (e.g., \cite{cohen1997photons}). In Sect. VI we validate this approximation. Furthermore, as we shall see later, $\Ham_{ii}'$ does not influence the natural motion of the plasmon modes.

We note that the Hamiltonian does not depend on the transverse component of $\xib \rp$. 

\section{Canonical Quantization}
The physical quantities of the system are quantized in a standard fashion (e.g. \cite{cohen1997photons,huttner1992quantization}) by enforcing the canonical commutation relations between the fields and their conjugates. To the canonically conjugate matter vector fields $\Pb \rp$  and $\xib \rp$ correspond, respectively, the Hermitian operators $\hP \rp$ and  $\hxi \rp$, and to the canonically conjugate radiation vector fields $\Pib$  and $\Ab$  correspond, respectively, the Hermitian operators $\hPi$  and $\hA$. We impose the equal time commutation relations:
\begin{subequations}
\begin{eqnarray}
\left[ \hP \rp, \hxi \rpp \right] &=& -i \hbar \I \delta \left( \rb - \rb' \right) \quad \rb, \rb' \in V, \\
\left[ \hPi \rp, \hA \rpp \right] &=& -i \hbar \I \delta^\perp \left( \rb - \rb' \right)  \quad \rb, \rb' \in V_\infty,
\end{eqnarray}
\end{subequations}
while all other commutators of the canonical variables vanish; $\I$ is the three-dimensional unit tensor and $\delta^\perp$ is the transverse delta function (e.g., \cite{cohen1997photons}).
\subsection{Normal mode expansion}
\label{sec:Expansion}
 We indicate with $\left\{ \mathbf{U}_m^\parallel \right\}$ an orthogonal basis of the functional space $\mathcal{U}^\parallel \left( V \right)$ of the longitudinal vector fields defined in $V$,  and with  $\left\{ \mathbf{U}_m^\perp \right\}$ an orthogonal basis of the functional space $\mathcal{U}^\perp \left( V \right)$ of the transverse vector fields defined in $V$, with $m$ running through discrete values (both these functional spaces have a discrete basis). 
 
 For diagonalizing the plasmon Hamiltonian term $\Ham_p$, we choose as basis for $ \mathcal{U}^\parallel \left( V \right)$ the set of the quasi-electrostatic plasmon modes of the nanoparticle \cite{mayergoyz2005electrostatic,mayergoyz2007dynamics}.
The normal mode $\Wl_m$ is given by
\begin{equation}
    \Wl_m \rp = - \nabla \oint_{\partial V} \frac{ w_m \rpp }{ 4 \pi\left| \rb - \rb' \right| } \dS' 
    \quad \text{in} \, V,
    \label{eq:potentialMain}
\end{equation}
where $w_m$ is solution of the eigenvalue problem 
\begin{equation}
 \label{eq:gammaMain}
     \mathcal{E}_s \left\{ w \right\} \rp = \frac{1}{\gamma} w \rp \, \text{with } {\bf r} \in \partial V,
     \end{equation}
     and
     \begin{equation}
    \mathcal{E}_s\left\{ w \right\} \rp = \oint_{\partial V} \frac{\left( \rb - \rb' \right)}{2\pi\left|  \rb - \rb' \right|^3} \cdot \n \rp w \rpp d^2 {\bf r}' \; \forall \mathbf{r} \in \partial V;
    \label{eq:OpMain}
\end{equation}
$\n$ is the normal to $\partial V$  pointing outward. The eigenvalues $\left\{ \gamma_m \right\}$ are real with $ \left| \gamma_m \right| > 1$. The eigenfunctions and the eigenvalue only depend on the shape of $V$, they do not depend on its size. The function $ w\rp $ represents the surface charge distribution of the quasi-electrostatic plasmon mode. For a detailed description of the properties of the operator $\mathcal{E}_s$ see Appendix  \ref{sec:A1}. 

As we shall see in the next Section, we do not need to chose explicitly a basis for $\mathcal{U}^\perp \left( V \right)$ because the transverse component of the electron gas motion does not influence the natural motion of the plasmon modes.

The field operators $\hP \rp$ and $\hxi \rp$ are represented by means of the following expansions:
\begin{subequations}
\begin{eqnarray}
\hP\rp &=& \displaystyle \sum_n \pl_n \frac{1}{\Vl_n} \Ul_n \rp + \sum_n \pt_n  \frac{1}{\Vt_n} \Ut_n \rp,  \\
\hxi \rp &=& \displaystyle\sum_n \ql_n \Ul_n \rp + \sum_n \qt_n \Ut_n \rp,
\end{eqnarray}
\end{subequations}
where $\left\{ \pl_m,\ql_m\right\}$ and $\left\{\pt_m,\qt_m \right\}$  are two sets of canonically conjugate Hermitian operators, and $\Vl_m = \left\| \Ul_m \right\|^2$, $\Vt_m = \left\| \Ut_m \right\|^2$. The vector fields $\Ul_m$  and $\Ut_m$  are dimensionless quantities, therefore $\Vl_m$  and $\Vt_m$ have the dimension of a volume, $\ql_m$   and $\qt_m$  have the dimension of a length,  $\pl_m$ and $\pt_m$ have the dimension of a linear momentum.

The functional space  $\mathcal{F} \left( V_\infty \right)$   of the canonically conjugate vector fields   $\Pib \rp$ and $\Ab \rp$ has a continuum basis. We indicate with $\left\{ {\bf f}_q \right\}$  an orthogonal basis of $\mathcal{F} \left( V_\infty \right)$  with $q$  running through a set of continuum values. In this paper we use the transverse plane waves as a basis for  $\mathcal{F} \left( V_\infty \right)$, namely,
\begin{equation}
\label{eq:planewave}
    \mathbf{f}_q \rp = \frac{1}{\left(2\pi\right)^{3/2}} \eps_{s,\mathbf{k}} e^{i \mathbf{k} \cdot \rb},
\end{equation}
where ${\bf k} \in \mathbb{R}^3$ is the propagation vector,  $\eps_{s,\mathbf{k}}$ is the polarization unit vector with $\eps_{s,\mathbf{k}} = \eps_{s,-\mathbf{k}}$ and $s=1,2$ (e.g., \cite{cohen1997photons}). 
The index $q$ is a multi-index corresponding to the pair of parameters $\mathbf{k}$ and $s$, $q = \left(s, \mathbf{k} \right)$.
The two polarization vectors are orthogonal between each other,  $\eps_{1,\mathbf{k}} \cdot \eps_{2, \mathbf{k}} = 0$,  and are both transverse to the propagation vector,  $\eps_{1,\mathbf{k}} \cdot \mathbf{k} = \eps_{2, \mathbf{k}} \cdot \mathbf{k}=0 $. The set of functions  $\left\{ \fb_q \right\}$ are orthonormal:
\begin{equation}
    \langle \fb_{q'}, \fb_q  \rangle = \delta_{s',s} \delta \left( \mathbf{k} - \mathbf{k}' \right).
\end{equation}
In the following, we denote $\delta_{s',s} \delta \left( \mathbf{k} - \mathbf{k}' \right) $ with $\delta_{q,q'}$  and  $\displaystyle\sum_{s=1}^2 \int_{\mathbb{R}^3} \left( \cdot \right) \dk$ with $\sum_q \left( \cdot \right)$.

The field operators $\hPi \rp$ and  $\hA \rp$ are represented through the following expansions:
\begin{subequations}
\begin{eqnarray}
     \hAb \rp &= \sum_q \hA_q \mathbf{f}_q \rp, \\
     \hPib \rp &= \sum_q \hPi_q \mathbf{f}_q^* \rp,
\end{eqnarray}
\end{subequations}
where $\hA_q$,  $\hPi_q$ are canonically conjugate operators. The operators $ \hAb$  and $\hPib$  are Hermitian, therefore $\hA_q^\dagger = \hA_{-q}$  and $\hPi_q^\dagger = \hPi_{-q} $.

The operators $\pl_m$, $\ql_m$ , $\pt_m$ , $\qt_m$ , $\hA_q$ , $\hPi_q$  verify the equal time commutation relations
\begin{subequations}
\begin{eqnarray}
    \left[ \pl_m, \ql_{m'} \right] &= - i \hbar \delta_{m,m'},  \\
    \left[ \pt_m , \qt_{m'} \right] &= - i \hbar \delta_{m,m'}, \\
    \left[ \hPi_q , \hA_{q'}^\dagger \right] &= - i \hbar \delta_{q,q'},
\end{eqnarray}
\end{subequations}
while all other commutators of these variables vanish. 

\subsection{Quantized Hamiltonian}

The quantized Hamiltonian is composed of four terms, plasmon $\hat{\Ham}_p$, radiation field $\hat{\Ham}_{em}$, plasmon – photon interaction $\hat{\Ham}_{i}$, and electron kinetic energy in the radiation field $\hat{\Ham}_{ii}'$  (we have disregarded the diamagnetic term  $\hat{\Ham}_{ii}''$)
\begin{equation}
    \hat{H} = \hat{\Ham}_p + \hat{\Ham}_{em} + \hat{\Ham}_{i} + \hat{\Ham}_{ii}'.
\end{equation}
The plasmon term $\hat{H}_p$ is given by:

\begin{equation}
    \hat{\Ham}_p = \sum_n \left( \frac{1}{ 2 \Ml_n} {\pl}_n{}^2 + \frac{\Ml_n \Omega_n^2}{2} {\ql}_n{}^2 \right),
\end{equation}
where $\Ml_m = \rho_0 \Vl_m$,
\begin{equation}
    \Omega_m = \omega_p \sqrt{ \frac{1}{2} \left( 1 - \frac{1}{\gamma_m} \right)},
\end{equation}
\begin{equation}
     \left\| \Ul_m \right\|^2= \frac{1}{4} \left( 1 - \frac{1}{\gamma_m^2} \right),
\end{equation}
$\Ul_m$ is given by Eq. \ref{eq:potentialMain} and $\gamma_m$ is the corresponding eigenvalue. The frequencies $\left\{ \Omega_m \right\}$ are the natural frequencies of the quasi-electrostatic surface plasmon modes of the nanoparticle. The first term in the expression of $\hat{\Ham}_p$ takes into account the contribution to the kinetic energy due to the irrotational component of the electron gas motion; the second term takes into account the contribution of the Coulomb potential energy. 

The expression of the radiation field term $\hat{\Ham}_{em}$  is
\begin{equation}
    \hat{\Ham}_{em} = \sum_q \left( \frac{1}{2 \varepsilon_0} \hPi_q^\dagger \hPi_q  + \frac{\varepsilon_0 \omega_q^2}{2} \hA_q^\dagger \hA_q \right),
\end{equation}
where $\omega_q^2 = c_0^2 k^2$, $c_0$  is the light velocity in vacuum. 
The expression of the plasmon-photon coupling term is
\begin{equation}
     \hat{\Ham}_i  = \frac{e}{m_e^*} \sum_{n,q} \frac{1}{\Vl_n} \langle \Ul_n, \fb_q \rangle \,  \pl_n \hA_q.
\end{equation}
The expression of the term $\Ham_{ii}'$ is
\begin{equation}
      \hHam_{ii}'  =  \sum_n \frac{1}{\Mt_n} {\pt}_n{}^2+ \frac{e}{m_e^*} \sum_{n,q} \frac{1}{\Vt_n}
    \langle \Ut_n, \mathbf{f}_q \rangle \,   {\pt}_n \hA_q,
\end{equation}
where $\Mt_m = \rho_0 \Vt_m$.

We now express the operators $\hA_q$, $\hPi_q$ and the operators $\pl_m$, $\ql_m$ by means of the standard boson operators $\ha_q$, $\ha_q^\dagger$ and $\hb_m$, $\hb_m^\dagger$,respectively. The operators $\ha_q$ and $\hb_m$ are defined as
\begin{subequations}
\begin{eqnarray}
    \ha_q &= \frac{1}{2} \left[ \sqrt{ \frac{2 \varepsilon_0 \omega_q}{\hbar}  } {\hA_q} + i \sqrt{\frac{2}{\varepsilon_0 \hbar \omega_q}} \hPi_q \right], \\
    \hb_m &= \frac{1}{2} \left[ \sqrt{ \frac{2 \Ml_m \Omega_m}{\hbar}} \ql_m + i \sqrt{\frac{2}{ \Ml_m\hbar \Omega_m}} \pl_m \right].
\end{eqnarray}
\end{subequations}
They verify the equal time commutation relations
\begin{subequations}
\begin{eqnarray}
     \left[ \ha_q, \, \ha^\dagger_{q'} \right] = \delta_{q,q'}, \\
     \left[ \hb_m, \, \hb^\dagger_{m'} \right] = \delta_{n,n'}, 
\end{eqnarray}
\end{subequations}
while all other commutators of the bosonic operators vanish. Note that $\ha_q^\dagger\ne\ha_-q$. The operators $\hb_m$ are dimensionless, while the operators $\ha_m$  have the dimensions of $\text{length}^{-3/2}$.
Then, the different terms of the Hamiltonian are expressed as:
\begin{subequations}
\begin{eqnarray}
    \hHam_p &=& \sum_n \hbar \Omega_n \left( \hb_n^\dagger \hb_n + \frac{1}{2} \right), \\
    \hHam_{em} &=& \sum_q \hbar \omega_q \left( \ha_q^\dagger \ha_q + \frac{1}{2} \right), \\
    \label{eq:hHint}
    \hHam_i &=& \sum_{n,q} V_q^n \left( \hb_n - \hb_n^\dagger \right)  \left( \ha_q + \ha_{-q}^\dagger \right), \\
    \hHam_{ii}' &=& \sum_n \frac{1}{2 \Mt_n} {\pt_n}{}^2  \\     
    &+& \frac{e}{m_e^*} \sum_{n,q} \frac{1}{\Vt_n} \langle \Ut_n, \fb_q \rangle \,  \pt_n \left( \ha_q + \ha^\dagger_{-q} \right),
    \nonumber
\end{eqnarray}
\end{subequations}
where
\begin{equation}
    V_q^n = i \frac{\hbar \omega_p}{2} \sqrt{\frac{\Omega_n}{\omega_q}} \frac{1}{\left\| \Ul_n \right\|} \langle \Ul_n , \fb_q\rangle.
\end{equation} 
We note that $\langle \Ul_m,\fb_q\rangle$ is the coefficient of the expansion of the transverse component of the vector field $\Ul_m$ in terms of transverse plane waves \ref{eq:planewave}. In Appendix A we show that to evaluate $\langle \Ul_m,\fb_q\rangle$ it is sufficient to compute a surface integral over the nanoparticle surface involving only the charge distribution of the quasi-electrostatic plasmon mode $\Ul_m$.

\subsection{Heisenberg equations}
We now use the Heisenberg picture, assuming the entire system to be initially in its ground state. Since the Hamiltonian does not depend explicitly on $\qt_m$, we have $\pt_m(t)=\pt_m(t=0)={\pt_m}{}^{\left(0\right)}$, where ${\pt_m}{}^{\left(0\right)}$ is the observable in the Schr\"odinger picture. Therefore, the expectation value at any time of ${\pt_m}{}$ is $0$. The time derivative of the observable $\qt_m$  is different from zero: it only depends on the time evolution of the vector potential observable.  Indeed, in the classical model $\mathbf{P}^\perp$  vanishes when the electron gas is initially at rest according to Eq. \ref{eq:Pd}. The component $\xib^\perp$  of the displacement field is evaluated once the vector potential $\mathbf{A}$ is known by using Eq. \ref{eq:Xd}. In the classical model $\left( \xib^\perp, \Pb^\perp \right)$ are not true state variables of the system: they do not influence the evolution of the longitudinal components $\left( {\xib}^\parallel, {\Pb}^\parallel \right)$.

The Heisenberg equations for $\hb_m \tp$ and $\ha_q \tp$ ($m=1,2,...$, and $q=(s,{\bf k})$ where $s = \eps_{1,\mathbf{k}}, \eps_{2,\mathbf{k}}$ and $ {\bf k}\in \mathbb{R}^3$) are
\begin{subequations}
\begin{eqnarray}
    \dot{{\hb}}_m &= \frac{1}{i \hbar} \left[ \hb_m \tp, \hHam \tp \right], \\
    \dot{{\ha}}_q &= \frac{1}{i \hbar} \left[ \ha_q \tp,  \hHam \tp \right],
\end{eqnarray}
\end{subequations}
where
\begin{subequations}
\begin{eqnarray}
    \left[\hb_m,  \hHam \right] &= \hbar \Omega_m\hb_m + \sum_q \left( V_q^m \ha_q - {V_q^m}^* \ha_q^\dagger \right),  \\
    \left[\ha_q,   \hHam \right] &= \hbar \omega_q \ha_q + \sum_n {V_q^n}^* \left( \hb_n - \hb_n^\dagger \right) + \hc_q,
\end{eqnarray}
\end{subequations}
and the constant operator $\hc_q$ is given by
\begin{equation}
    \hc_q = \frac{e}{m_e^*} \sum_n \frac{1}{\Vt_n} \langle \fb_q , \Ut_n \rangle \, {\pt_n} {}^{\left(0 \right)}.
\end{equation}
Therefore, the equations of motion for $\dot{\hb}_m$  and $\dot{\ha}_m $ are
\begin{subequations}
\begin{eqnarray}
 \label{eq:dotb}
    \dot{\hb}_m &+& i \Omega_m \hb_m + \frac{i}{\hbar} \sum_q \left( V_q^m \ha_q - {V_q^m}^* \ha_q^\dagger \right) = 0, \\
     \label{eq:dota}
    \dot{\ha}_q &+& i \omega_m \ha_q + \frac{i}{\hbar} \sum_n  {V_q^n}^* \left( \hat{b}_n - \hat{b}_n^\dagger  \right) = -\frac{i}{\hbar} \hc_q.
\end{eqnarray}
\end{subequations}

\section{Natural frequencies of the dressed plasmon modes}

We now consider the evolution of the  expectation values of the operators $\left\{ \hb_n \right\}$ and $\left\{ \ha_n \right\}$ in the Heisenberg picture, $\beta_m  \left( t \right) = \langle \hb_m \tp \rangle$ and  $\alpha_q   \left( t \right) = \langle \ha_q \tp \rangle$. The average of $\hc_q$ over the ground state is 0 at any time. From Eqs. \ref{eq:dotb} and \ref{eq:dota}  we obtain
\begin{subequations}
\begin{eqnarray}
    \label{eq:dBeta}
    \dot{\beta}_m &+& i \Omega_m \beta_m + \frac{i}{\hbar} \sum_q \left( V_q^m \alpha_q - {V_q^m}^* \alpha_q^* \right) = 0,\\
    \label{eq:dotAlpha}
    \dot{\alpha}_q &+& i \omega_q \alpha_q + \frac{i}{\hbar} \sum_n {V_q^n}^* \left( \beta_n - \beta_n^* \right) = 0.
\end{eqnarray}
\end{subequations}
Equation \ref{eq:dotAlpha} gives:
\begin{equation}
 \label{eq:alfa}
    \alpha_q \left( t \right) = -\frac{i}{\hbar} \sum_n V_q^{n}{}^* \int_{0}^{\infty} h_q \left( t - \tau \right) \left[ \beta_n \taup - \beta_n^* \taup \right] d\tau,
\end{equation}
where
\begin{equation}
    h_q \tp = \theta \tp e^{- i \omega_q t},
\end{equation}
and $\theta \tp$ is the Heaviside function. By substituting \ref{eq:alfa} into \ref{eq:dBeta}, we obtain the following system of integro-differential equations:
\begin{widetext}
\begin{equation}
 \label{eq:dBeta1}
    \dot{\beta}_m + i \Omega_m \beta_m + \frac{1}{\hbar^2} \sum_{n,q}  \int_{0}^{\infty} \left[ V_q^{m} V_q^{n}{}^* h_q \left( t - \tau \right) - V_q^{m}{}^* V_q^{n} h_q^* \left( t - \tau \right)  \right] \left[ \beta_n \taup - \beta_n^* \taup \right] d\tau = 0, m = 1,2,... .
\end{equation}
\end{widetext} 
To determine the free evolution we solve this system with the initial conditions: $\beta_1 \left( 0^+ \right) = 1 $, $\beta_m \left( 0^+ \right) = 0$ for $m \ne 1$; $\beta_2 \left( 0^+ \right) = 1 $, $\beta_m \left( 0^+ \right) = 0$ for $m \ne 2$; and so on. These problems are solved by applying the Laplace transform, $F\left( s \right) = \int_{0}^{\infty} f\left(t\right)e^{-st} dt$.
Denoting the Laplace transform of $\beta_m \tp$  by  $\BetaS \left( s \right)$ and of $\beta_m^*\tp$  by  $\BetaSS \left( s \right)$, Eqs. \ref{eq:dBeta1} become in the Laplace domain:
\begin{equation}
 \label{eq:dBeta3}
    \left( s + i \Omega_m \right) \BetaS_m + \Omega_m \sum_n R_{mn} \left( \BetaS_n - \BetaSS_n \right) = \delta_{mk},
\end{equation}
where
\begin{multline}
    R_{mn} \left( s \right) = \frac{1}{\hbar^2 \Omega_m} \sum_q \left( \frac{V_q^m {V_q^n}^*}{s + i \omega_q} 
    - \frac{{V_q^m}^* V_q^n }{s- i \omega_q} \right),
\end{multline}
and $m,k=1,2,\ldots$. The Laplace transform converges for $Re(s)>0$. 
 
  \begin{figure}
    \centering
    \includegraphics[width=0.8\columnwidth]{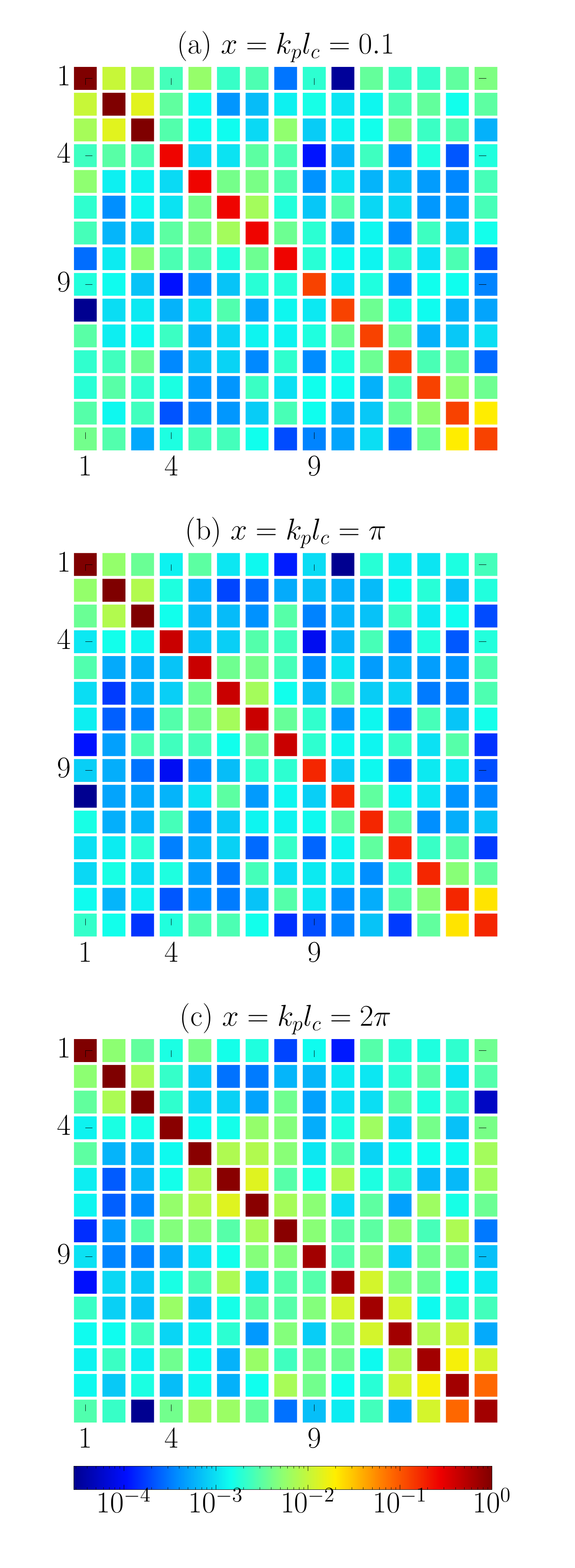}
\caption{ Magnitude of the matrix elements $R_{mn}$, computed in $s = - i \Omega_k + 0^+ $, for a sphere with different values of $k_p l_c$ where $l_c$  is the radius. In each panel, each occurrences have been normalized to $\displaystyle\max_{m}{\left| {R}_{mm} \right|}$. The coupling among the electric dipoles ($m,n = 1,2,\ldots 3$), quadrupoles $\left(m, n = 4 \ldots 8\right)$, and octupoles $\left( m, n = 9 \ldots 15 \right)$ have been considered.}
    \label{fig:Matrix}
\end{figure}

In the system \ref{eq:dBeta3} the term $\Omega_m \sum_n R_{mn} \left( \BetaS_n - \BetaSS_n \right)$ accounts for the plasmon - photon coupling. The self-coupling term ${\Omega_m} R_{mm} \left( \BetaS_m - \BetaSS_m \right)$ is only responsible for the coupling between the $m$-th quasi-electrostatic plasmon mode and the photons, while the remaining terms determine an additional coupling among different quasi-electrostatic plasmon modes mediated by the radiation. Nevertheless, we  found that the mutual coupling coefficients $\bar{R}_{mn}$ are negligible with respect to the self-coupling coefficients  ${R}_{mm}$  in a wide range of nanoparticle sizes because $\left| R_{mn}\right| \ll \left| R_{mm}\right|$ for $n \ne m$ and any  $m$. This is due to the orthogonality of the quasi-electrostatic plasmon modes. In Fig. \ref{fig:Matrix}, for example, we represent $\left|\bar{R}_{mn}\right|/\displaystyle\max_{m} \left|\bar{R}_{mm} \right|$ for the first 15 modes of a sphere, including  three  degenerate dipole  modes ($m, n = 1, 2, 3$),  five  degenerate  quadrupole  modes ($m, n = 4, \ldots, 8$),  and seven degenerate octupole modes ($m, n = 9, \ldots, 15$), evaluated at $s=-i \Omega_k+0^+$ according to the pole approximation technique (e.g. \cite{barnett2002methods}) that we apply later. We consider three values of the size parameter $x= k_p l_c$, where $k_p = \omega_p / c_0$ and $l_c$ here is the sphere radius. In all the investigated cases, the amplitudes of the mutual coupling coefficients are smaller than the amplitudes of the self-coupling coefficients (of at least two orders of magnitude). We have found similar results for all the shapes and arrangements we have investigated (for example, in the bow-tie antenna the off diagonal terms $R_{mn}$ are at least three order of magnitude smaller than the  terms $R_{mm}$ on the diagonal). Therefore, the self-coupling coefficients certainly prevail over those of mutual coupling. This numerical evidence prompted us to  disregard the off diagonal terms in the system of equations \ref{eq:dBeta3}. Thus, we obtain for $m = 1,2, ...$ the following system of two equations:
\begin{equation}
    \label{eq:uncoupled}
    \begin{pmatrix}
    \BetaS_m  \\  
    \BetaSS_m
    \end{pmatrix}
    +M_m
    \begin{pmatrix}
    \BetaS_m \\  
    \BetaSS_m
    \end{pmatrix}
    =
    \begin{pmatrix}
    1  \\  
    1
    \end{pmatrix}
\end{equation}
where 
\begin{equation}
    M_m \left( s \right) = \Omega_m
    \begin{pmatrix}
    i + R_{mm}     & -R_{mm} \\
    R_{mm} & -\left(i + R_{mm}\right) 
    \end{pmatrix}.
\end{equation}
This is a central result of our analysis. The plasmon - photon coupling gives rise to dressed plasmon modes, but the radiation does not induce a relevant coupling among the different quasi-electrostatic plasmon modes for particles of sizes up to the plasma wavelength. The effects of the interaction of the $m^{th}$ quasi-electrostatic plasmon mode with the photons, represented by the term $ \Omega_m R_{mm}(s) \left( \BetaS_m - {\BetaSS_m} \right)$, cause mainly both a frequency shift with respect to the quasi-electrostatic frequency, and an exponentially decay in time of the plasmon amplitude.

We solve equation \ref{eq:uncoupled} by using the pole approximation technique (e.g. \cite{barnett2002methods}), under which ${R_{mm}\left(s\right)}$ is approximated by its value at $s = \left(- i \Omega_m + 0^+ \right)$, which we denote with ${\rho_m}$,
\begin{equation}
    \rho_m = - \frac{i}{\hbar^2 \Omega_m} \sum_q \left| V_q^m \right|^2 \left( \frac{1}{\Omega_m + \omega_q} + \frac{1}{\omega_q - \Omega_m - i 0^+} \right).
\end{equation}
Since (e.g. \cite{heitler1984quantum}))
\begin{equation}
    \lim_{\epsilon \rightarrow 0} \frac{1}{x- i \epsilon} = \mathcal{P} \frac{1}{x} + i \pi \delta,
\end{equation}
we have:
%\begin{widetext}

\begin{multline}
 \label{eq:rhoPole}
    \rho_m = \frac{1}{\hbar^2 \Omega_m} \sum_q \left| V_q^m \right|^2 \left[ \pi \delta \left( \omega_q - \Omega_m \right) \right. \\ \left. - \mathcal{P} \frac{i}{\omega_q - \Omega_m} - \frac{i}{\Omega_m + \omega_q} \right],
\end{multline}
%\end{widetext}
where $\mathcal{P}$ denotes the Cauchy principal value. Therefore, the complex natural frequency of the $m^{th}$ plasmon mode, indicated with $\lambda_m$, is the eigenvalue with negative real part of the matrix
\begin{equation}
    M_m \left( - i \Omega_m + 0^+ \right) = \Omega_m
    \begin{pmatrix}
    i + \rho_m     & -\rho_m \\
    \rho_m & -\left(i + \rho_m\right) 
    \end{pmatrix}.
\end{equation}
It is given by
\begin{equation}
\lambda_m =-i \Omega_m \sqrt{1- 2 i \rho_m}.
\label{eq:LambdaPole}
\end{equation}
The eigenvalue with positive real part is instead a spurious solution introduced by the pole approximation, which has to be disregarded in the study of the positive time dynamics. Then, the radiative decay rate $\Gamma_m$ and the frequency shift $\Delta \Omega_m$ of the $m^{th}$  plasmonic mode, with respect to the quasistatic resonance frequency $\Omega_m$, are given by
\begin{subequations}
\begin{eqnarray}
     \label{eq:Gamma}
     \Gamma_m &=&  2\text{Re} \left\{ \lambda_m \right\}, \\
     \label{eq:Delta}
     \Delta \Omega_m &=&- \text{Im} \left\{ \lambda_m \right\} - \Omega_m.
\end{eqnarray}
\end{subequations}
Equation \ref{eq:LambdaPole} represents the main finding of the present manuscript. Combined with Eqs. \ref{eq:Gamma} and \ref{eq:Delta} it provides a closed-form full-wave expression for the decay rate and for the frequency shift of plasmon modes, which only depend on the surface charge distribution of the quasi-electrostatic plasmon mode.
r
We could have taken an alternative path for the derivation of the natural frequencies of the dressed plasmon modes, by introducing a further, more restrictive, approximation in the Hamiltonian $\hHam$. Specifically, by applying the rotating wave approximation to the term $\hHam_i$ of the Hamiltonian in Eq. \ref{eq:hHint},  and following the same steps of the previous section, including neglecting the coupling among different plasmon modes, we obtain the following natural frequency of the $m^{th}$ mode:
\begin{equation}
    \lambda_m^{\text{RWA}} = -i \Omega_m \left( 1 - i \rho_m^\text{RWA} \right),
    \label{eq:LambdaRWA}
\end{equation}
where
\begin{equation}
    \rho_m^{\text{RWA}} = \frac{1}{\hbar^2 \Omega_m} \sum_q \left| V_q^m \right|^2 \left[ \pi \delta \left( \omega_q - \Omega_m \right) - \mathcal{P} \frac{i}{\omega_q - \Omega_m}  \right].
    \label{eq:rhoRWA}
\end{equation}
The radiative decay rate and the frequency shift in the RWA are then obtained using Eqs. \ref{eq:Gamma} and \ref{eq:Delta}, provided that we replace $\lambda_m$ with $\lambda_m^{\text{RWA}}$. We point out that, even in the limit $\left|{\rho_m}\right| \ll 1$, the natural frequency given by Eqs. \ref{eq:rhoPole}, \ref{eq:LambdaPole} differs from the the natural frequency $\lambda_m^{\text{RWA}}$ predicted by the RWA through Eqs. \ref{eq:LambdaRWA} and \ref{eq:rhoRWA}, because they differ of the imaginary quantity $-\frac{i}{\Omega_m+\omega_q}$. Based on these considerations, we espect that, in the limit of very small particles, the RWA returns the correct decay rate, but not the correct frequency shift.

In the next Section we validate both Eqs. \ref{eq:LambdaPole}, \ref{eq:LambdaRWA} against the Mie theory for a metal nano-sphere of radius comparable to the plasma wavelength.

\section{Radiative decay rate and frequency shift of metal nanoparticles}

In this Section,  we investigate the radiative decay rate $\Gamma_k$ and the frequency shift  $\Delta \Omega_k$ of the $k^{th}$ plasmonic mode with respect to the quasistatic resonance frequency of metal nanoparticle as a function of the size parameter $x= k_p l_c$, where $k_p = \omega_p / c_0$ and $l_c$ is a characteristic size of the nanoparticle. We analyze nanoparticles with different shapes and spatial arrangements.

\begin{figure}
    \centering
    \includegraphics[width=\columnwidth]{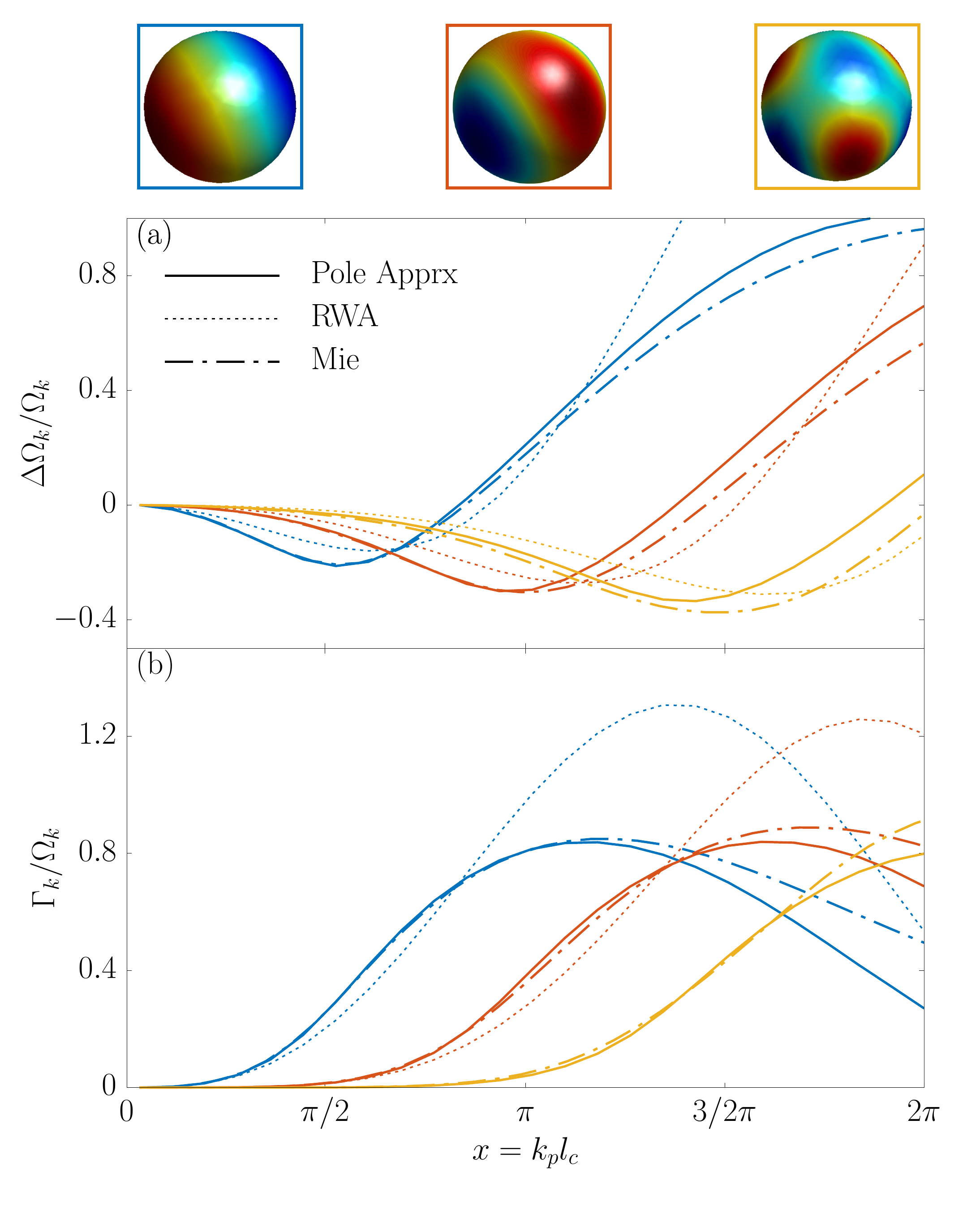}
\caption{Frequency shift $\Delta \Omega_k$ with respect to the quasistatic resonance frequency (a) and decay rate $\Gamma_k$ (b) normalized to the quasi-electrostatic resonance frequency $\Omega_k$ of the dipole (blue), quadrupole (red), octupole (yellow lines) modes of a sphere as a function of the size parameter $x = k_p l_c$ where $l_c$ is the radius $R$. Three different approaches have been used: the poles of the Mie coefficients (dash-dots), the pole approximation (continuous line), and the rotating wave approximation RWA (dots). The surface charge density distributions of the modes are shown above, enclosed in a box whose color matches the color of the corresponding curve.}
    \label{fig:Sphere}
\end{figure}

\begin{figure}
    \centering
    \includegraphics[width=\columnwidth]{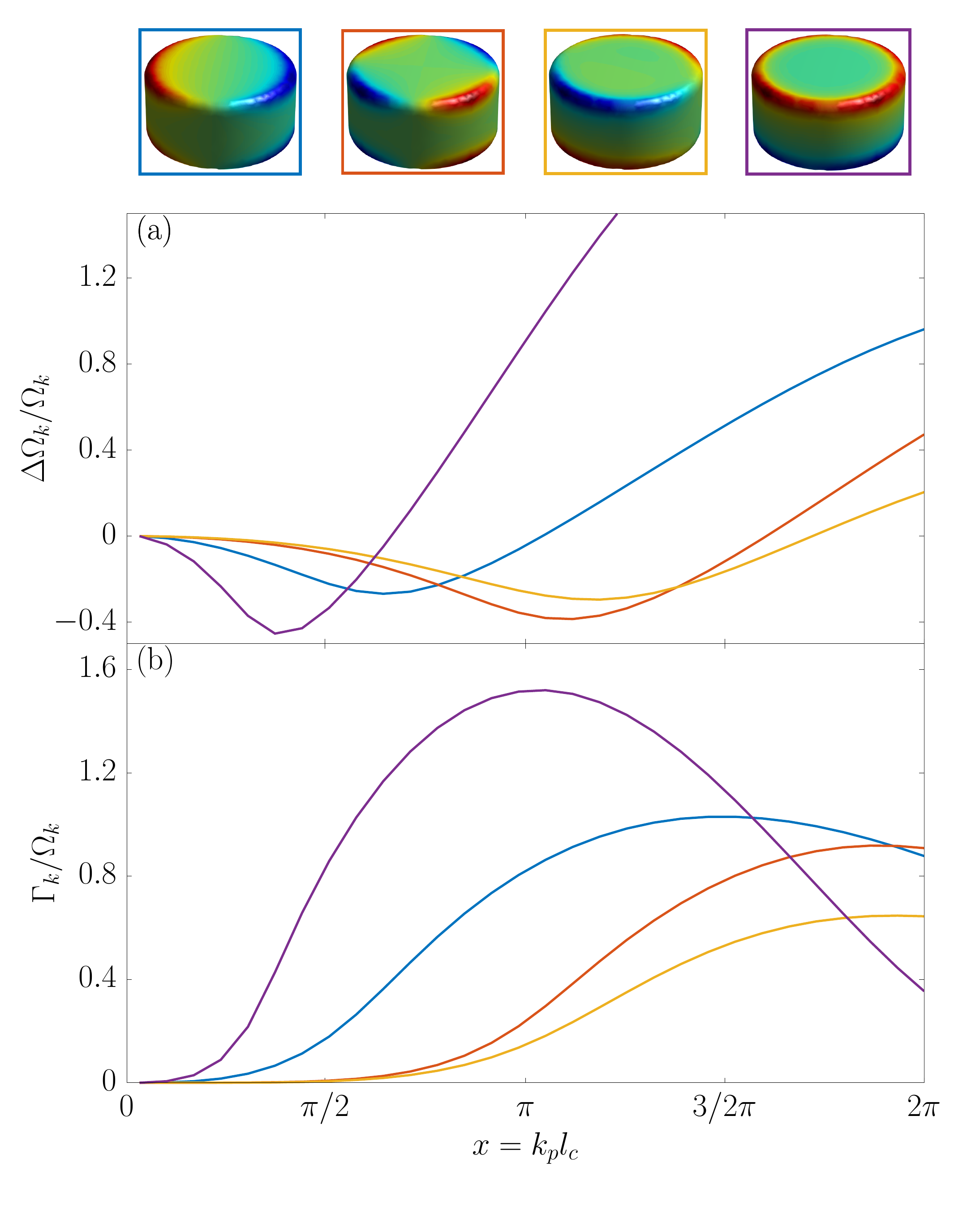}
\caption{Frequency shift $\Delta \Omega_k$  with respect to the quasistatic resonance frequency (a) and decay rate $\Gamma_k$ (b) normalized to the quasi-electrostatic resonance frequency $\Omega_k$ of the first four non-degenerate modes of a finite-size cylinder of radius $R=l_c$ and height $H=R$. The surface charge density distributions of the modes are shown above, enclosed in a box whose color matches the color of the corresponding curve.}
    \label{fig:Cyl}
\end{figure}

\begin{figure}
    \centering
    \includegraphics[width=\columnwidth]{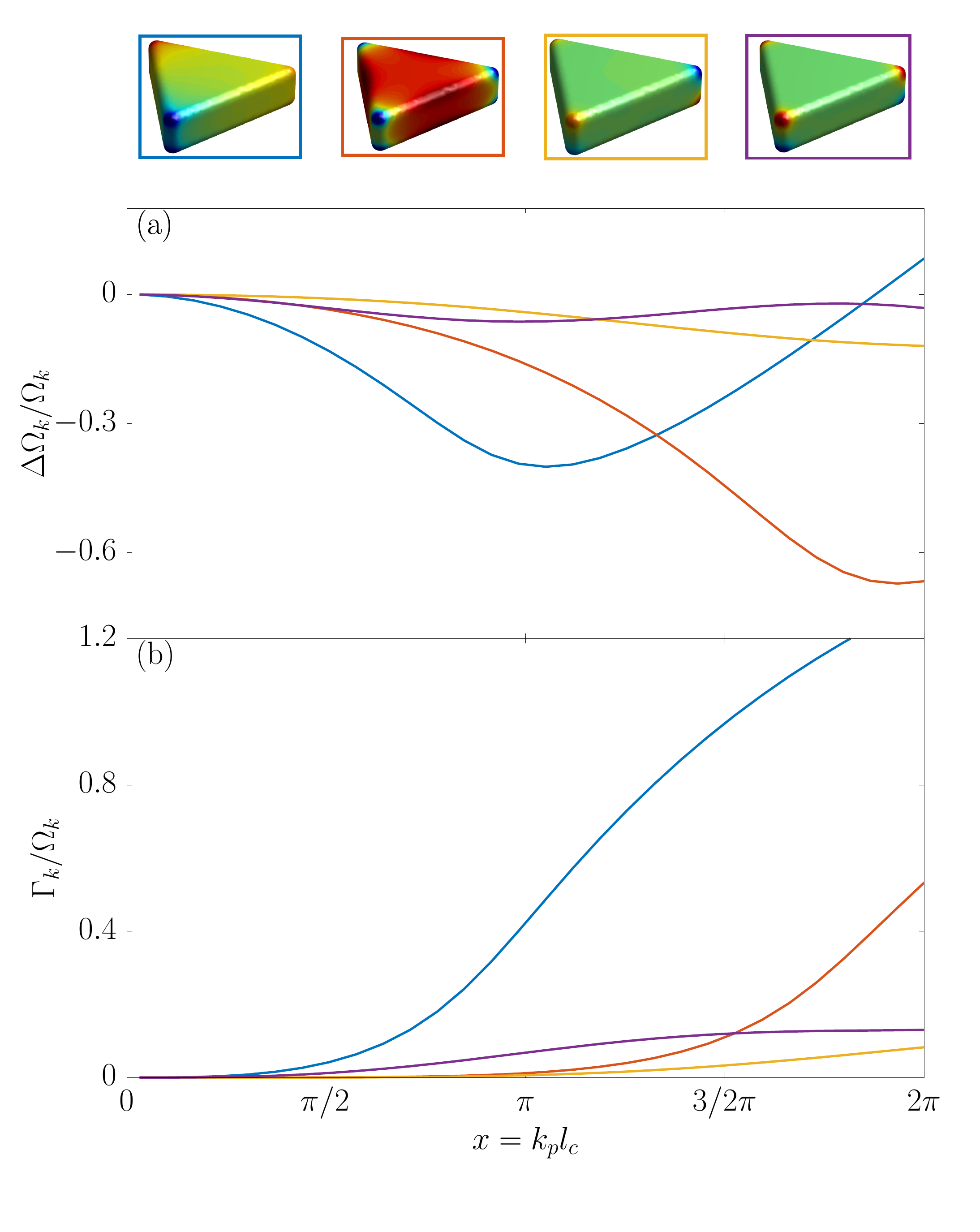}
\caption{Frequency shift $\Delta \Omega_k$ {\color{blue} with respect to the quasistatic resonance frequency} (a) and decay rate $\Gamma_k$ (b) normalized to the quasi-electrostatic resonance frequency $\Omega_k$ of a triangular prism of basis edge $2l_c$ and height $l_c/2$, as a function of the size parameter $x = k_p l_c$. The surface charge density distributions of the modes are shown above, enclosed in a box whose color matches the color of the corresponding curve.}
    \label{fig:Tri}
\end{figure}

First, aiming at a validation of our approach, we study a sphere where an analytic solution, i.e. the Mie Theory \cite{hergert2012mie}, exists. We consider the first $15$ modes (including three degenerate dipole modes, five degenerate quadrupole modes, and seven degenerate octupole plasmonic modes). In Fig. \ref{fig:Sphere} we show the frequency shift $\Delta \Omega_k / \Omega_k$ (a) and the radiative decay rate $\Gamma_k / \Omega_k$  (b) of the dipole (blue lines), quadrupole (red lines) and octupole (yellow lines) plasmonic modes of a sphere as a function of $x$. Three different approaches are compared: the Mie-theory (dot-dash), the pole approximation (continuous line) computed from Eq. \ref{eq:LambdaPole}, and the rotating wave approximation (RWA) (dots) computed from Eq. \ref{eq:LambdaRWA}. The quasi-electrostatic surface charge density distributions associated to the dipole, quadrupole and octupole modes are shown above, enclosed in a box whose color matches the color of the corresponding curves. 

The results obtained by using the pole approximation are in overall good agreement with the results obtained by the Mie theory in the whole range $\left[ 0, 2 \pi \right]$.  We recall that i) we neglected the diamagnetic term \ref{eq:dia} in the Hamiltonian and ii) we applied the pole approximation technique to solve Eq. \ref{eq:uncoupled}.  The good agreement we found shows that these approximations are well founded in the range $\left[ 0 , 2 \pi \right]$. This is a core aspect for what concerns the modelling of devices and experiments, since this range includes all the metal nanoparticle sizes that are of practical use in nanophotonics. For instance, contextualizing our results to existing materials: for a gold sphere with $\omega_p \approx 6.79$ T rad/s \cite{Maier:03}, the upper limit of the investigated range $x= 2 \pi$ corresponds to  $R=270$ nm (diameter of $540$ nm). It is very uncommon to encounter larger particles in the applications.

On the contrary, the rotating wave approximation, even if it is able to {\it qualitatively} capture the shape of the curve of frequency shift and decay rate, gives numerical predictions that appreciably deviate from the poles of the Mie coefficients.  Specifically, in the limit of very small particle $x \rightarrow 0$, while the pole approximation and the Mie Theory perfectly agree,  the rotating wave approximation only captures the slope of the decay rate, but not the slope of the frequency shift. The reason for this discrepancy has already been pointed out at the end of the previous section.

Having validated our model, we now use it to investigate the behavior of the decay rate and of the frequency shift of nanoparticles with different shapes and spatial arrangements. For the sphere, as the size parameter increases, starting from $x=0$,  all the modes first undergo a relative red shift with respect to their quasi-electrostatic resonance frequency and then a blue shift,  as shown in Fig. \ref{fig:Sphere} (a). Lower multipolar modes reach the maximum relative red shift for smaller size parameters than higher multipoles. Similarly, decay rates of dipole, quadrupole and octupole modes, increase for small size parameters, reach a point of maximum, and then decreases,  as shown in  Fig.  \ref{fig:Sphere}(b). This behavior has been already pointed out in Ref.\cite{kolwas2013damping}.

By using now only the pole approximation of Eq. \ref{eq:LambdaPole}, we analyze the frequency shift and the decay rate of the plasmon modes of cylinders and triangular prisms. In Fig. \ref{fig:Cyl} we show the frequency shift and the decay rate of the first four non-degenerate modes of a finite-size cylinder with $l_c = R = H$, whose rounded edges have a radius of curvature $0.1 l_c$ ($R$ is the radius of the cylinder and $H$ is the height) as the size parameter varies.  In particular, the horizontal dipole mode (blue curve) of the cylinder behaves similarly to the dipole mode of the sphere, while the vertical dipole (purple curve) features a larger decay rate and a larger frequency shift. The second and the third mode are quadrupoles and behave similarly to the corresponding mode of a sphere.

Next, we investigate the triangular prism of edge $2 l_c$, height $0.5 l_c$ shown in Fig. \ref{fig:Tri}, whose rounded edges have a radius of curvature $0.1 l_c$. The first mode is a horizontal dipole (blue curve) and features an higher radiative decay rate if compared to the dipole of a sphere. Among the remaining modes we also recognize a quadrupole ($3$-rd mode) and a vertical dipole  ($4$-rd mode).

\begin{figure*}
    \centering
    \includegraphics[width=\textwidth]{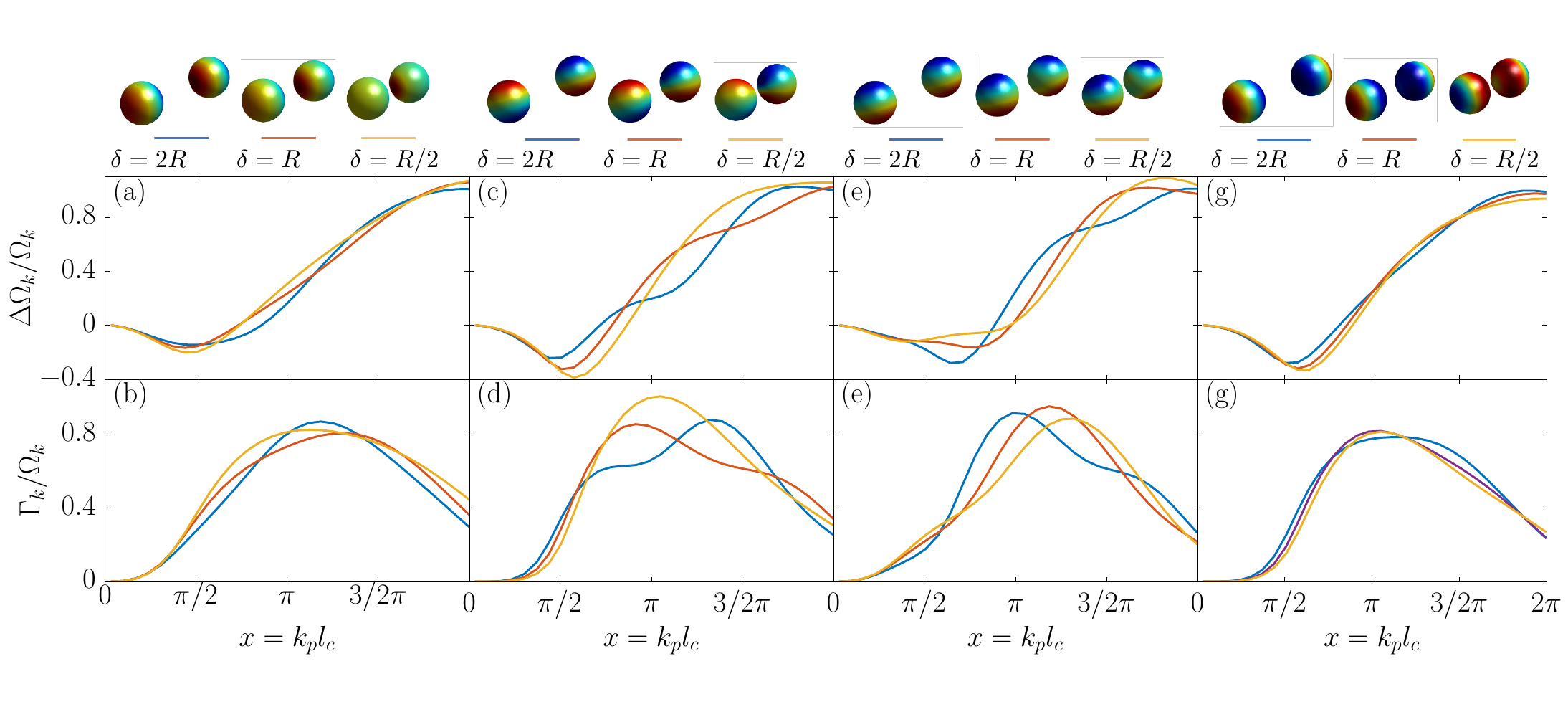}
\caption{Frequency shift $\Delta \Omega_k$  with respect to the quasistatic resonance frequency and decay rate $\Gamma_k$ normalized to the quasi-static resonant frequency $\Omega_k$ of the first four modes of a sphere dimer as a function of the size parameter $x = k_p l_c$. Different edge-edge gap sizes have been considered, namely $\delta=2l_c$, $\delta=l_c$, $\delta=l_c/2$. The charge density distributions of the plasmon modes are shown above the corresponding panels.}
    \label{fig:SphereDimer}
\end{figure*}

\begin{figure*}
    \centering
    \includegraphics[width=\textwidth]{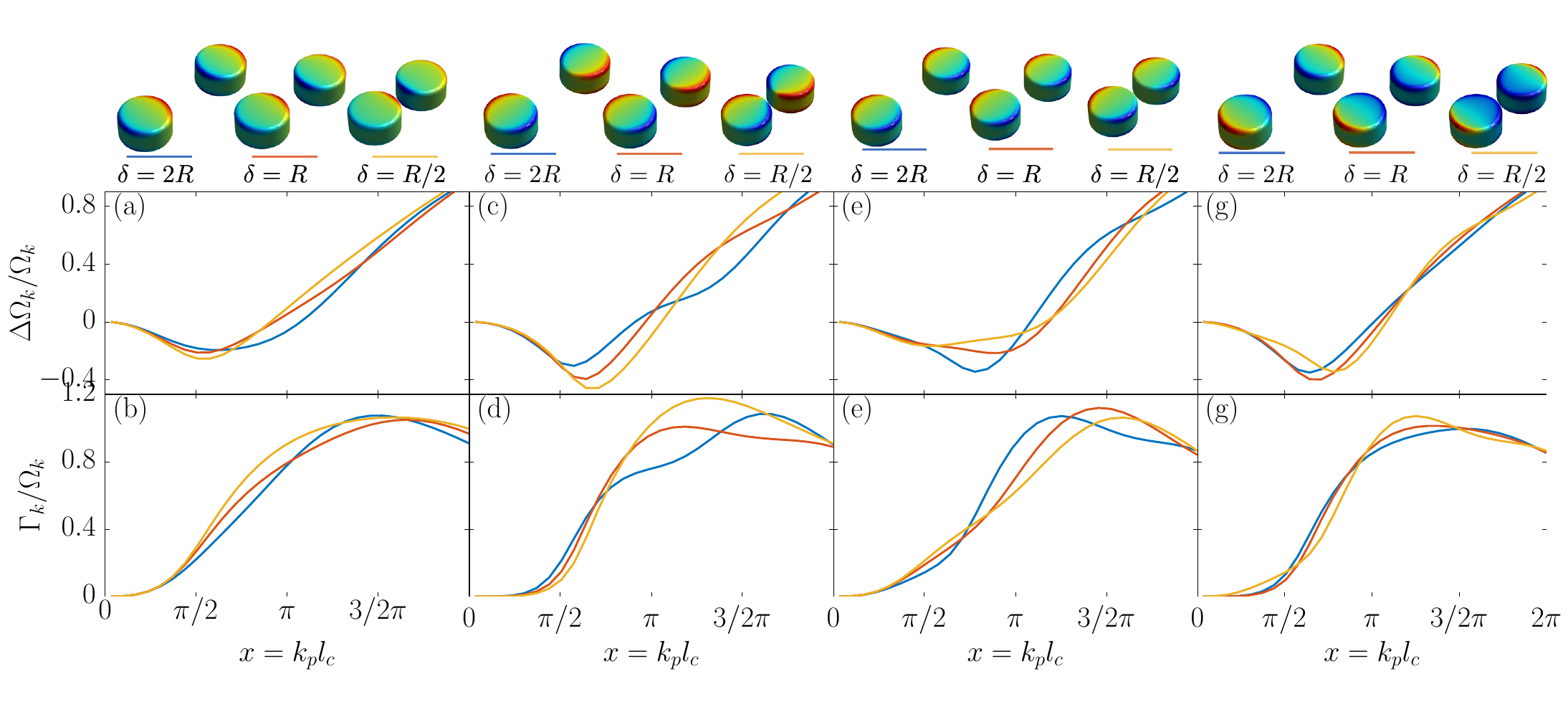}
\caption{Frequency shift $\Delta \Omega_k$   with respect to the quasistatic resonance frequency and decay rate $\Gamma_k$, normalized to the quasi-electrostatic resonant frequency $\Omega_k$, of the first four modes of a sphere dimer as a function of the size parameter $x = k_p l_c$. Different edge-edge gap sizes have been considered, namely $\delta=2l_c$, $\delta=l_c$, $\delta=l_c/2$. The charge density distributions of the plasmon modes are shown above the corresponding panels.}
    \label{fig:CylDimer}
\end{figure*}

\begin{figure*}
    \centering
    \includegraphics[width=\textwidth]{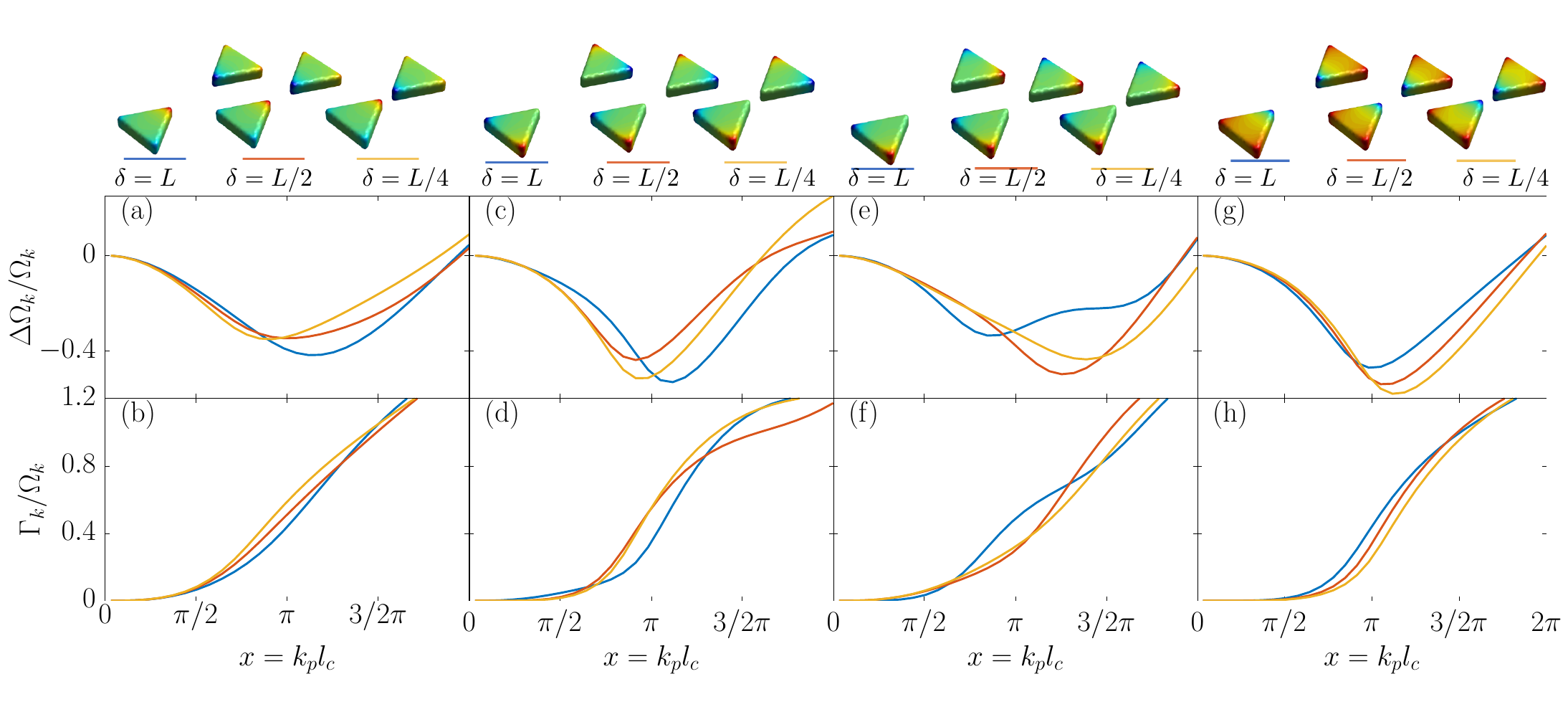}
\caption{Frequency shift $\Delta \Omega_k$ with respect to the quasistatic resonance frequency and decay rate $\Gamma_k$, normalized to the quasi-electrostatic resonant frequency $\Omega_k$, of the first four modes of a bow-tie antenna as a function of the size parameter $x = k_p l_c$. Different edge-edge  gap sizes have been considered, namely $\delta=2l_c$, $\delta=l_c$, $\delta=l_c/2$. The charge density distributions of the plasmon modes are shown above the corresponding panels.}
    \label{fig:BowTie}
\end{figure*}

We now employ the nanoparticles we investigated so far as building blocks of dimers. Specifically, we investigate the frequency shift and the decay rate of the first four non-degenerate plasmon modes, for a dimer of spheres in Fig. \ref{fig:SphereDimer}, for a dimer of cylinders in Fig. \ref{fig:CylDimer}, and for a bow-tie antenna in Fig. \ref{fig:BowTie}. These modes originate from the hybridization of the corresponding dipole modes of the isolated building blocks. Different edge-edge gaps have been considered, namely $\delta=2l_c$, $\delta=l_c$, $\delta=l_c/2$.
In each panel we monitor a specific dimer-mode as the edge-edge gaps is decreased, showing how the surface charge density distribution, the frequency shift, and the decay rate are affected by the change in the gap size. Despite the difference among the different shapes of the building blocks, we found similar trends of the corresponding modes.
In particular, in Figs. \ref{fig:SphereDimer}, \ref{fig:CylDimer}, and \ref{fig:BowTie} the $1$st and the $4$th mode are the {\it bonding} and {\it anti-bonding} longitudinal (with respect to the dimer's axis) dipole mode, respectively. Similarly the $2$nd and $3$rd mode are the antibonding and bonding transverse dipole modes.

Overall both $\Omega_k$ and $\Gamma_k$ respectively resemble the shift and decay rate exhibited by the dipole mode of the corresponding isolated building blocks. Nevertheless, several shoulders arise from the radiative coupling between the charges of the two constituent nanoparticles.

\section{Conclusions}  
We  introduced a time-domain model of the natural motion of the dressed plasmon modes in metal nanoparticles, which arise from the interaction between  plasmons and photons.  The plasmon - photon coupling is described in the framework of the Hopfield model for the plasmon oscillations, where the plasmon oscillations are represented by a harmonic matter field linearly coupled to the electromagnetic radiation.  The main effects of plasmon - photon coupling are a frequency shift with respect to the quasi-electrostatic frequency, and an exponentially decay in time of the plasmom mode amplitude, while we found the coupling among different quasi-electrostatic plasmon modes to be negligible for nanoparticles of sizes up to the plasma wavelength.  We also found the diamagnetic term in the Hamiltonian to be negligible.

By solving the equations governing the expectation values of the plasmon creation and annihilation operators, we have derived new closed-form full-wave expressions for the frequency shift and for the decay rate of the plasmon natural modes. 

We found the frequency shift and the decay rate of the dressed plasmons by using the pole approximation, obtaining good quantitative agreement with the Mie theory for spheres of radius up to the plasma wavelength. Instead, we found that rotating wave approximation fails to describe their natural evolution even for very small particles.

Then, we studied the decay rate and the frequency shift of the plasmon modes in isolated nanoparticles and dimers of different shapes, as their size increases up to the plasma wavelength. In all the investigated cases, as the size increases, the resonance frequency of the natural plasmon modes first undergoes a relative red shift with respect to its corresponding quasi-electrostatic position, which is then followed by a blue shift. Similarly, the decay rate increases for small sizes, reaches a maximum and then decreases.

The developed approach leads, as expected from the correspondence principle, to the same outcome obtained of the classical Hamiltonian equations. Nevertheless, it constitutes a first step toward a full-wave time-domain quantum framework based on a modal analysis for describing the interaction between a quantum emitter, an arbitrary shaped metal nanoparticle and, the electromagnetic radiation.

\appendix

% APPENDIX 1
\section{Longitudinal normal modes}
\label{sec:A1}
In this Appendix we introduce the orthogonal basis to represent the functional space $\mathcal{U}^\parallel \left( V \right)$ of the solenoidal and irrotational vector fields defined on the bounded region $V$ that diagonalize the plasmon term $\Ham_p$ of the Hamiltonian of the system. Eventually, we provide two simple formulas to compute the coefficients $\left\| \Ul_m \right\|^2$  and $\langle \Ul_m , \fb_q \rangle_V$.
%% A.1
\subsection{Single and double layer scalar potentials}
We consider the single layer scalar potential
\begin{equation}
    \varphi \rp = \frac{1}{4 \pi} \oint_{\partial V} \frac{ w \rpp }{ \left| \rb - \rb' \right| } \dS' \quad {\bf r} \in V_\infty
\end{equation}
that is generated in the whole space by a free-standing layer of charge with surface density $w \rp$ located on the boundary  $\partial V$ of $V$. We introduce the integral operator 
\begin{equation}
    \mathcal{E}_s \left\{ w \right\} \rp = \frac{1}{2\pi} \oint_{\partial V} \frac{\left( \rb - \rb' \right)}{\left|  \rb - \rb' \right|^3} \cdot \n \rp w \rpp d^2 {\bf r}' \; \forall \mathbf{r} \in \partial V,
\end{equation}
where $\n$ is the normal to $\partial V$  pointing outward. Then, the normal derivative to $\partial V$ of $\varphi \rp $ at $\partial V$ is given by
\begin{equation}
-\frac{\partial \varphi^\pm}{\partial n} = \mp \frac{1}{2} w + \frac{1}{2} \mathcal{E}_s \left\{ w \right\} \text{on }\partial V,
\end{equation}
where $\varphi^\pm$  are the restriction of $\varphi \rp$  in  $V$ and $V_\infty/V$, respectively.

We now also consider the scalar potential $\phi \rp$ generated in the whole space by a free-standing double layer of charges located on $\partial V$ with surface density $\nu \rp$,
\begin{equation}
    \phi \rp = \frac{1}{4 \pi} \oint_{\partial V} \frac{\left( \rb - \rb ' \right)}{\left| \rb -\rb' \right|^3} \n \rpp v \rpp \dS' \, \text{with} \rb \in V_\infty.
\end{equation}
We also introduce the integral operator $\mathcal{E}_s^\dagger$ dual to the integral operator $\mathcal{E}_s \left\{ w \right\}$,
\begin{equation}
    \mathcal{E}_s^\dagger \left\{ \nu \right\} \rp = \frac{1}{2\pi} \int_{\partial V} \frac{\left( \rb - \rb ' \right)}{\left|  \rb - \rb ' \right|^3} \cdot \n \rp \nu \rpp d^2 {\bf r}' \qquad {\bf r} \in \partial V.
\end{equation}
Then, the scalar potential $\phi \rp$ is given by
\begin{equation}
    \phi^\pm = \pm \frac{1}{2} \nu + \frac{1}{2} \mathcal{E}_s^\dagger \left\{ v \right\},
\end{equation}
where $\phi^\pm$ denote the restriction of $\phi \rp $  in  $V$ and $V_\infty/V$, respectively.

%% A.2
\subsection{Auxiliary eigenvalue problems}

We now introduce the eigenvalue problem:
\begin{eqnarray}
 \label{eq:gamma}
     \mathcal{E}_s \left\{ w \right\} \rp = \frac{1}{\gamma} w \rp \, \text{with } {\bf r} \in \partial V.
\end{eqnarray}
The operator $\mathcal{E}_s$  is symmetric but not Hermitian, and its spectrum has the following properties ( [41, 42]):
\begin{enumerate}[i]
    \item The set of eigenvalues $\left\{ \gamma_m \right\}$  and the set of eigenfunctions $\left\{ w_m \right\}$  are infinite countable;
    \item The eigenvalues and the eigenfunctions are real, and  $ \left| \gamma_m \right| \ge1$;
    \item The eigenvalue and the eigenfunctions depend on the shape of $V$ but not on its sizes;
    \item The normal derivative of the scalar potential $\phi_m$ generated by the surface charge density $w_m$ is given by
    \begin{eqnarray}
     \label{eq:8}
       \frac{\partial \varphi_m^\pm}{\partial n} = \left( \pm 1 - \frac{1}{\gamma_m} \right) \frac{w_m}{2} \, \text{on } V.
    \end{eqnarray}
\end{enumerate}
The operator $\mathcal{E}_s^\dagger$ is the adjoint of the operator $\mathcal{E}_s$. It  has the following properties:

\begin{enumerate}[i]
\setcounter{enumi}{4}
\item Its spectrum coincides with the spectrum of the integral operator $\mathcal{E}_s$ except for the eigenvalue $\gamma=1$; 
\item The eigenfunction $v_n$ of $\mathcal{E}_s^\dagger$ associated to the eigenvalue $\gamma_n$ and  the eigenfunction $w_m$  of $\mathcal{E}_s$ associated to the eigenvalue $\gamma_m$ for $n \ne m$ are orthogonal, namely
\begin{eqnarray}
\label{eq:9}
  \oint_{\partial V} w_m\rp v_n\rp \dS = 0\, \text{ for $n \ne m$ }.
\end{eqnarray}

\item The scalar potential $\phi_m \rp$  generated by the double layer surface density $v_m$  is given by
\begin{equation}
\label{eq:10}
   \phi_m^\pm = \left( \pm 1 + \frac{1}{\gamma_m} \right) \frac{v_m}{2} \quad \text{on} \; \partial V.
\end{equation}
\item Under proper normalization,  the scalar potential $\varphi_n \rp$ generated by the surface charge density  $w_m$, and the scalar potential $\phi_n \rp$  generated by the double layer surface density $v_n$  are equal,
\begin{eqnarray}
 \label{eq:11}
    \varphi_n = \phi_n \quad \text{in} \, V_\infty.
\end{eqnarray}
\end{enumerate}
%%%%%%%%%%%%
% A.2
%%%%%%%%%%%%

\subsection{Basis for linear functional space $\mathcal{U}^\parallel \left( V \right)$  and diagonalization of the plasmon Hamiltonian}

We now introduce the vector field $\Wl_m \rp$ generated in the whole space by the surface charge with density $w_m \rp$,
\begin{equation}
    \Wl_m \rp = - \nabla \varphi_m \quad \text{in} \, V_\infty.
    \label{eq:potential}
\end{equation}
The vector field $\Wl_m$  is irrotational everywhere in $V_\infty$, it is solenoidal both in $V$  and in $V_0 = V_\infty \backslash V$ , the normal components to $\partial V$  are different from zero, and they are discontinuous.

 If we indicate with $\Wl_m  {}^{\left( + \right)} \cdot \n $  the normal component of $\Wl_m$ on the outer page of $\partial V$  and with $\Wl_m {}^{\left( - \right)} \cdot \n $ the normal component of $\Wl_m$ on the inner page we have
\begin{eqnarray}
 \label{eq:wm}
   \left. {\Wl_m}^{\left(\pm \right)} \right|_{\partial V} \cdot \n = \pm \frac{1}{2} w_m + \frac{1}{2} \mathcal{E}_s \left\{ w_m \right\} \text{on} \,\partial V. \quad
\end{eqnarray}
The vector fields $\left\{ \mathbf{U}_m^\parallel \right\}$  have the following properties (\cite{fredkin2003resonant,mayergoyz2005electrostatic}):
\begin{enumerate}[i]
    \setcounter{enumi}{8}
    \item $\langle  \Wl_{m'}, \Wl_{m} \rangle_V = \delta_{m',m} \left\| \Wl_m \right\|^2_{V} $;  \\
    \item $\langle  \Wl_{m'}, \Wl_{m} \rangle_{V_0} = \delta_{m',m} \left\| \Wl_m \right\|^2_{V_0} $;  \\
    \item $\left( \gamma_m + 1 \right) \left\| \Wl_m\right\|^2_{V} = \left( \gamma_m - 1 \right) \left\| \Wl_m\right\|^2_{V_0}$;
    \item $\left\| \Wl_m \right\|^2_{V_\infty} =  \frac{2 \gamma_m}{ \gamma_m - 1} \left\| \Wl_m \right\|^2_V$;
    \item the set of the vector fields $\left\{ \mathbf{U}_m^\parallel \right\}$  is complete in the functional linear space of the irrotational and solenoidal functions defined on $V$ and $V_0$ with discontinuous normal components on $\partial V$.
\end{enumerate}

We now use as basis for the functional space $\mathcal{U}^\parallel$  introduced in Section \ref{sec:Expansion} the restriction to the domain $V$  of the set of vector fields $\left\{ \Wl_m \right\}$. This choice allows us to diagonalize the plasmon Hamiltonian term  $\Ham_{p}$. By expanding $\hxi^\parallel$ and $\hP^\parallel$ as 
\begin{eqnarray}
    \hxi^\parallel = \sum_m \ql_m \Ul_m \rp,
\end{eqnarray}
\begin{eqnarray}
    \hP^\parallel = \sum_m \pl_m \Ul_m \rp,
\end{eqnarray}
by using \ref{eq:gamma}, \ref{eq:wm} and the relation
\begin{equation}
\oint_{\partial V} \oint_{\partial V} \frac{w_m \rp w_{m'} \rpp}{4 \pi \left| \rb - \rb' \right|} \dS \dS' =  \left\| \Wl_m \right\|^2_{V_\infty} \delta_{m,m'},
\end{equation}
we obtain for the plasmon Hamiltonian term
\begin{equation}
    \hat{\Ham}_p = \sum_n \left( \frac{1}{ 2 \Ml_n} {\pl_n}{}^2 + \frac{\Ml_n \Omega_n^2}{2} \ql_n{}^2 \right),
\end{equation}
where $\Ml_m = \rho_0 \Vl_m$, $\Vl_m = \left\| \Ul_m \right\|^2$, and
\begin{equation}
    \Omega_m = \omega_p \sqrt{ \frac{1}{2} \left( 1 - \frac{1}{\gamma_m} \right)}.
\end{equation}

%%%%%%%
%   A3  
%%%%%%%

\subsection{Evaluation of $\left\| \Ul_m \right\|^2$ and $\langle \Ul_m , \fb_q \rangle_V$}

We now evaluate the quantity $\left\| \Ul_m \right\|^2$ by using the Gauss theorem. We have:
\begin{equation}
     \left\| \Ul_m \right\|^2= \oint_{\partial V} \varphi_m^{\left(-\right)} \frac{ \partial \varphi_m^{\left(-\right)}}{\partial n} \dS.
\end{equation}
By using \ref{eq:8}, \ref{eq:11}, \ref{eq:10} and \ref{eq:9} we obtain:

\begin{equation}
     \left\| \Ul_m \right\|^2= \frac{1}{4} \left( 1 - \frac{1}{\gamma_m^2} \right).
\end{equation}

We now evaluate the quantity $\langle \Ul_m, \fb_q \rangle_V$ by using again the Gauss theorem. By using Eq. \ref{eq:potential} and $\nabla \cdot \fb_q=0$ in  $V$, we have:
\begin{equation}
 \langle \Ul_m , \fb_q \rangle_V = - \oint_{\partial V} \fb \rp \cdot \n \rp \varphi \rp \dS.
\end{equation}

\end{document}